\documentclass[prl,floats,twocolumn]{revtex4}
\usepackage[final]{graphicx}
\DeclareGraphicsExtensions{.eps,.ps,.pdf}
\usepackage{amsmath,amssymb,bbold,bm}
\newcommand\vex[1]{\mathbf{#1}}

\def\sslash#1{\setbox0=\hbox{$#1$}			
   \dimen0=\wd0                                		
   \setbox1=\hbox{/} \dimen1=\wd1  	 		
   \ifdim\dimen0>\dimen1                               	
      \rlap{\hbox to \dimen0{\hfil / \hfil}} 	  	
      #1                                      			
   \else                                        			
      \rlap{\hbox to \dimen1{\hfil$#1$\hfil}}   		
      \hbox{/} 	                              			
   \fi}   
\def\slash#1{\hbox{$#1$\kern-0.35em\raise0.1ex\hbox{/}}}

\begin{document}

\title{Signatures of surface states in bismuth at high magnetic fields}

\author{Babak Seradjeh}
\email{babaks@illinois.edu}
\affiliation{Department of Physics, University of Illinois, 1110 West Green St, Urbana 61801, USA.}
\author{Jiansheng Wu}
\affiliation{Department of Physics, University of Illinois, 1110 West Green St, Urbana 61801, USA.}
\author{Philip Phillips}
\affiliation{Department of Physics, University of Illinois, 1110 West Green St, Urbana 61801, USA.}

\begin{abstract}
Electrons in a metal subject to magnetic field commonly exhibit oscillatory behavior as the field strength varies, with a period set by the area of quantized electronic orbits. Recent experiments on elemental bismuth have revealed oscillations for fields above 9 tesla that do not follow this simple dependence and have been interpreted as a signature of electron fractionalization in the bulk. We argue instead that a simple explanation in terms of the surface states of bismuth exists when additional features of the experiment are included. These surface electrons are known to have significant spin-orbit interaction. We show the observed oscillations are in quantitative agreement with  the surface theory, which we propose to test by studying the effect of the Zeeman coupling in higher fields, dependence on the field orientation, and the thickness of the samples.
\end{abstract}

\maketitle

Elemental bismuth has recently attracted renewed interest due to the experimental observation by Behnia~\emph{et al.} of anomalous quantum oscillations at high magnetic fields $\gtrsim9$~T along the trigonal crystal axis~\cite{BehBalKop07a}. Plateau-like features in the Hall resistivity in this range of fields were taken as an indication of bulk electron fractionalization, in a manner reminiscent of the fractional quantum Hall (FQH) effect in two dimensions (2d). A second experiment by Li~\emph{et al.} found evidence for a new phase of electrons at such high fields~\cite{LiCheHor08a}. Some theoretical work has followed~\cite{AliBal08a,ShaMik09a}; however, the anomalies~\cite{BehBalKop07a} remain unresolved.

The idea of fractionalization in a bulk material is quite intriguing, especially in an isotropic semimetal such as bismuth. To this date, the only known realizations of fractionally charged particles are solitonic excitations of conducting polymers~\cite{Hee03a}, FQH state of 2d semiconductor heterostructures~\cite{Lau83b,TsuStoGos82a}, and confined quarks of quantum electrodynamics. Highly anisotropic relatives of the \emph{integer} quantum Hall effect are found in layered semiconductor heterostrctures~\cite{StoEisGos86a} and Bechgaard salts~\cite{BalKriWil95a,McKHanSch95a} but no FQH state yet. The key experimental signature of FQH states is a plateau in the sideways resistivity at fractional values of $h/e^2$ and a concomitant vanishing of the longitudinal resistivity. While the stability of FQH states derives from Coulomb interaction and weak disorder, their existence owes to the essential role of topology in 2d. Theoretical extensions of fractionalization in 3d rely either on an anisotropic structure or on background topological structures such as solitons in a relativistic Dirac Hamiltonian. It is not clear that such anisotropy as in the first scenario occurs in bismuth and although electrons in bismuth are known to have a Dirac-like dispersion, it is not clear what field would provide the topological background in the second one. Furthermore, the longitudinal resistivity in bismuth is found to be rather featureless at such field orientations~\cite{BehBalKop07a,FauYanShe09a}. The torque magnetometry~\cite{LiCheHor08a}, which is a sensitive bulk measurement, does not show these anomalies either, casting further doubt on a FQH scenario. We believe based on these considerations an alternative explanation of the observed anomalies is strongly favored.

Indeed we shall show that they can be explained in a simple fashion by the states confined to the (111) surface of bismuth. In particular, a distinct indexing of Landau levels according to the surface states with a period of oscillation $\approx0.016$~T$^{-1}$ explains the experimental data. The surface theory predicts: (1) the existence of additional features that might have already been observed and an additional peak at higher fields $\sim 60$~T corresponding to the surface quantum limit; (2) a distinct 
dependence of the field at the peaks on the angle $\theta$ of a tilting field relative to the trigonal axis. The magnetic fields so far studied fall below the surface quantum limit. However, the Zeeman coupling of surface electrons could reduce the quantum limit down to $40$--$50$~T. These features should allow for a falsification of the surface-state theory of the anomalous peaks.

Quantum oscillations are ubiquitous in metals. 
As the field changes, the Landau levels cross the Fermi energy and depopulate their electrons at certain fields $B_n^{-1} = (n+\gamma) \Delta B^{-1}$, where $0<\gamma<1$ is a correction arising from the quantum mechanical nature of the electronic orbit ($\gamma=1/2$ for free electrons) and
\begin{equation}\label{eq:OLK}
\Delta B^{-1} = \frac{(2\pi)^2}{\Phi_0 S_F}.
\end{equation}
Here, $S_F$ is an extremal area of the Fermi surface perpendicular to the field and $\Phi_0=hc/e$ is the unit flux quantum.
The coincidence of the Fermi and Landau energies results in enhanced or singular contributions to most electronic properties; hence they show an oscillatory structure with $B^{-1}$ with the period of oscillations given by the Onsager relation, Eq.~(\ref{eq:OLK}).

Due to its low carrier density and long mean-free path, bismuth exhibits periodic oscillations in its susceptibility and resistivity starting at fields below $1$~T~\cite{Ede76a,SteBab55b}. Using Eq.~(\ref{eq:OLK}) and the known Fermi surfaces of electron- and hole-like carriers in bismuth, one finds that $9$~T corresponds to depopulating all but the lowest Landau level for holes, followed closely by that of electrons, the so-called quantum limit. A recent study of the Landau levels of electrons and holes in bismuth, including the effects of the linear dispersion of electrons, accounts for most of the peaks observed below this field~\cite{ShaMik09a} with distinct dependence on the field orientation. The anomalous peaks~\cite{BehBalKop07a} do not follow Eq.~(\ref{eq:OLK}).

Bismuth also has a variety of interesting electronic states confined to its surfaces~\cite{Hof06a}. The (111) surface perpendicular to the trigonal axis has been studied in depth using ARPES and other techniques and found to host a number of carriers equivalent to a $1~\mu$m-thick sample of the bulk. The samples in the experiment~\cite{BehBalKop07a} have a thickness of $0.8$~mm; so, despite their higher relative carrier density one expects the Nernst signal of the surface to be weaker than that of the bulk by a factor of at least 1000. This is certainly true for fields below the quantum limit. However, above the quantum limit, the strong peaks from the bulk will be absent and the surface signal could be detected much more easily. The surface states in bismuth and related materials have also gained attention in relation to ``topological insulators''~\cite{XiaWraQia08a}. The important ingredient in such physics is the significant spin-orbit interaction (SOI)  that causes a large energy splitting in the surface states and leads to a single spin state per momentum. In bismuth the role of SOI has long been appreciated and has been shown to be the cause for the existence of hole Fermi surfaces~\cite{Hof06a}. At the (111) surface, the SOI results in six elongated hole Fermi pockets arranged around an electron pocket centered at the $\overline\Gamma$ point in the hexagonal Brillouin zone. 

\begin{figure}[tb]
\begin{center}
\includegraphics[height=4in]{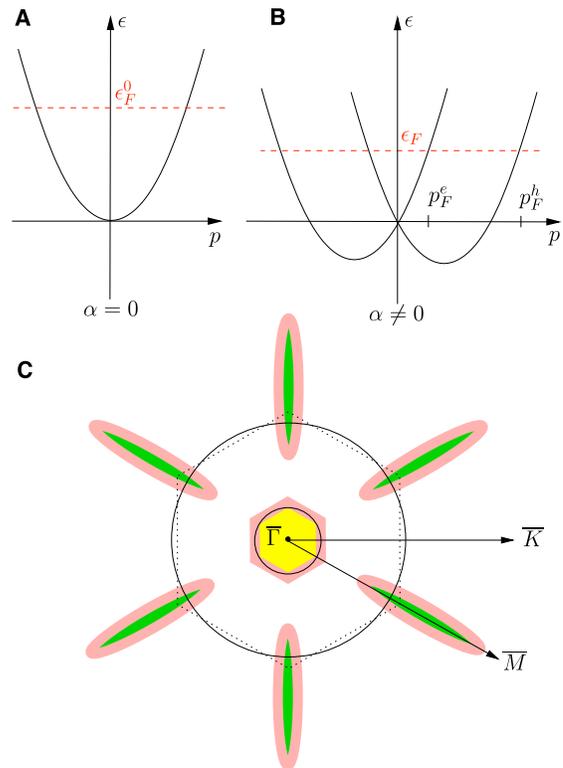}
\end{center}
\caption{(color online) The spectrum of the surface hamiltonian (A) without and (B) without spin-obit interaction. The Fermi energies are marked, as are the Fermi wavevectors for the small and large Fermi pockets. (C) Sketch of Fermi pockets in the surface Brillouin zone (to scale). The two circles are from our model. The central (yellow) hexagon and the outward (green) ellipses are from ARPES measurements~\cite{Hof06a}. The thickness of the lines (pink) represents the experimental resolution. The dashed hexagon indicates the symmetry of the pockets and the Brillouin zone.}
\label{fig:spec}
\end{figure}

In order to study the effects of the Zeeman coupling and tilting fields we model the surface states of Bi(111) by a low-energy continuum hamiltonian
\begin{equation}\label{eq:H}
H = \frac{{\vex p}^2}{2m} + \alpha \boldsymbol{\sigma}\cdot(\vex p\times \hat{\vex z}),
\end{equation}
where $\vex p$ is the momentum operator, $m$ is the effective mass of electrons, $\boldsymbol{\sigma}=(\sigma_x,\sigma_y)$ are spin Pauli matrices, and the last term is the Rashba-type SOI with strength $\alpha$. The trigonal axis is taken to be along the $\hat{\vex z}$ direction. The spectrum of Eq.~(\ref{eq:H}) is shown in Fig.~\ref{fig:spec}A-B and consists of a small and a large Fermi pocket. There is a single spin state for each momentum $\vex p$ given by a spin along $\vex p\times\hat{\vex z}$. In this model both of these pockets are electron-like, which is an artifact of the model. In reality, for higher momenta away from the $\overline\Gamma$ point, the lattice effects become important. This results in a bending of the bands, giving rise to hole-like pockets. The lattice anisotropy further reduces the symmetry creating disjoint oblongular hole pockets. Based on angle-resolved photoemission (ARPES) data~\cite{Hof06a}, we set $m=0.45m_e$ ($m_e$ is the real electron mass), $\alpha=1.2$~eV\AA\ and $\epsilon_F=68$~meV. These parameters reproduce the radius of the small electron pocket $p_F^e=-m\alpha+\sqrt{(m\alpha)^2+2m\epsilon_F}=0.043$~\AA$^{-1}$ (see Fig.~\ref{fig:spec}C) within the experimental resolution ($\sim0.023$~\AA$^{-1}$) and give a Fermi velocity $v_F=p_F^e/m+\alpha=1.9$~eV\AA\ in good agreement with the data. The lower band crosses the Fermi energy at $p_F^h=p_F^e+2m\alpha=0.185$~\AA$^{-1}$ corresponding to one end of the hole pocket.

In 2d, the period of oscillations is also given by Eq.~(\ref{eq:OLK}) but with $S_F$ now simply being the area of the Fermi pocket. We shall see this explicitly in our model.
The orbital effect of the magnetic field is found by the minimal coupling $\vex p \to \vex p-\frac ec\vex A$ in Eq.~(\ref{eq:H}), where $\vex A$ is the vector potential. The Zeeman energy is $H_{\mathrm{Z}}=-g\mu_B\sigma_z B$ for a field along $\hat{\vex z}$, where $\mu_B=e\hbar/(2m_ec)
$ is the Bohr magneton. 
The experimental determination of the $g$-factor usually relies on matching the Zeeman energy splitting with the Landau level spacing, and thus requires a knowledge of the spectrum itself. In the following we shall take $g$ as a fitting parameter which allows its determination from our theory in a tilting field.

The spectrum in the magnetic field can be found exactly. Let us first define
\begin{equation}
\Pi=\boldsymbol{\sigma}\cdot\left(\vex p - \frac ec \vex A\right)\times\hat{\vex z} - \frac{g_{\mathrm{eff}}\mu_B}{\alpha} \sigma_z B,
\end{equation}
with $g_{\mathrm{eff}}=g-m_e/m$.
Then by using the the commutation relations $[p_x-\frac ec A_x,p_y-\frac ec A_y]=i(\hbar e/c)B$ and the identity
\begin{equation}\label{eq:Pi2}
\Pi^2=\left(\vex p - \frac ec \vex A\right)^2 - \frac{\hbar e}{c}\sigma_z B + \left(\frac{g_{\mathrm{eff}}\mu_B}{\alpha} B\right)^2 + \hbar \frac{g_{\mathrm{eff}}\mu_B}{\alpha}\boldsymbol{\sigma}\cdot\boldsymbol{\nabla} B ,
\end{equation}
we find
\begin{equation}\label{eq:HPi}
H=\frac{\Pi^2}{2m}+\alpha\Pi-\frac1{2m}\left(\frac{g_{\mathrm{eff}}\mu_B}{\alpha} B\right)^2.
\end{equation}
We have assumed the spatial variation of the field to be negligible, which is justified when the last term in Eq.~(\ref{eq:Pi2}) is small compared to the Landau level spacing $\sim\hbar\omega_c$. This yields $\delta B/B\ll L/\lambda$ where $\delta B$ is the spatial variation of the field over a characteristic length $L$, and $\lambda=g_{\mathrm{eff}} \hbar/(4m_e\alpha)=g_{\mathrm{eff}}\times1.6$~\AA. This is in fact a much less strict condition than the one needed for obtaining Landau levels when $g_{\mathrm{eff}}=0$, namely, $\delta B/B_n<n^{-1}$ for the $n$th Landau level.

The spectrum of $\Pi$ is the same as that of Dirac electrons in a magnetic field with velocity $\alpha$ and Zeeman coupling $g_{\mathrm{eff}}$. This is known in the context of the quantum Hall effect in graphene, and is given by
\begin{equation}
E_{n,s} = \pm\sqrt{2m\hbar\omega_c(n+s)+\left(\frac{g_{\mathrm{eff}}\mu_B}{\alpha} B\right)^2}.
\end{equation}
Here $n=0,1,\cdots$ and $s=0,1$ are Landau level indices and $\omega_c=e B/(mc)$ is the cyclotron frequency.  From Eq.~(\ref{eq:HPi}), we have
\begin{equation}\label{eq:spec}
\epsilon_{n,s}^\pm = \hbar\omega_c(n+s)\pm\sqrt{2m\alpha^2\hbar\omega_c(n+s)+(g_{\mathrm{eff}}\mu_B B)^2}.
\end{equation}
The $\pm$ correspond to small ($-$) and large ($+$) Fermi pockets. Note that the level $n, s=1$ are degenerate with $n+1, s=0$. The lowest Landau level $n=s=0$ is non-degenrate.

Let us first neglect the effective Zeeman coupling by setting $g_{\mathrm{eff}}=0$. Then by setting $\epsilon_{n,s}=\epsilon_F$ we can derive Eq.~(\ref{eq:OLK}) with $S_F=S_F^e=\pi (p_F^e)^2$ for the small electron pocket and similarly for the large one. Consequently, the fundmamental period of oscillations in the surface theory is then found to be
\begin{equation}
\Delta B^{-1}=0.016~\mathrm{T}^{-1}.
\end{equation}
 The Zeeman term changes the linear dependence of $1/B_n$ on the Landau level index. This can be understood by formally neglecting the orbital coupling and keeping only the Zeeman coupling $H_{\mathrm{Z}}$. Then one finds two bands $\epsilon(\vex p, \sigma)=\frac{\vex p^2}{2m}+\sigma \sqrt{\alpha^2\vex p^2+(g_{\mathrm{eff}}\mu_B B)^2}$, where $\sigma=\pm1$ is the sign of the spin projected along $\alpha\vex p\times\hat{\vex z}+g_{\mathrm{eff}}\mu_BB\hat{\vex z}$. As a result, the small electron pocket shrinks and the large Fermi pocket expands by the same area $\Delta S_F=\frac{p_F^h-p_F^e}{p_F^h+p_F^e}\left(\frac{g_{\mathrm{eff}}\mu_B}{\alpha} B\right)^2$. The Fermi energy is unchanged. With our choice of parameters, $\Delta S_F/S_F^e=7.4\times10^{-7}(g_{\mathrm{eff}} B/\mathrm{T})^2\approx 10\%$ when $g_{\mathrm{eff}} B=370$~T. The field-dependent area of the Fermi pocket then leads to a nonuniform Landau level spacing and the depopulation of Landau levels at lower fields. In particular the surface quantum limit $B_1(g_{\mathrm{eff}})$ 
is a decreasing function of $g_{\mathrm{eff}}$. The deviation from the linear dependence should be observable in fields $\gtrsim 40$--$50$~T for a value of $g_{\mathrm{eff}}\sim10$. 

\begin{figure}[tb]
\begin{center}
\includegraphics[height=4.2in]{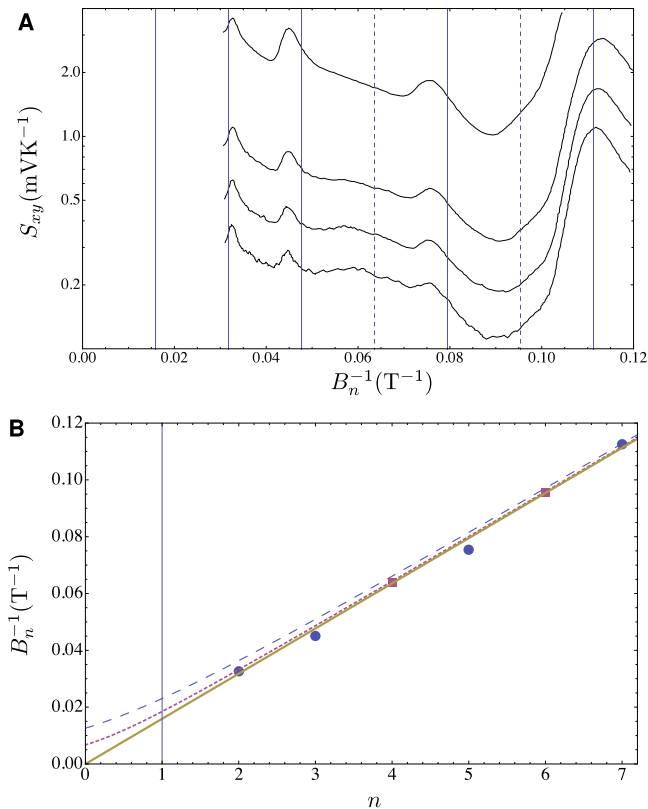}
\end{center}
\caption{(color online) (A) The Nernst signal of Ref.~\onlinecite{BehBalKop07a} (digitized) with the surface Landau levels ($n=1,\cdots 7$) marked. The dashed lines are $n=4$ and $6$. (B) Fits of the inverse field at the peaks from Eq.~(\ref{eq:spec}). The peaks identified in Ref.~\onlinecite{BehBalKop07a} are shown by circles. The straight line is obtained for $g_{\mathrm{eff}}=0$ for which $B^{-1}_4$ and $B^{-1}_6$ are shown by squares. The dotted and dashed curves are for $g_{\mathrm{eff}}=8$ and $15$, respectively. The quantum limit is obtained at the intersection with the vertical line.}
\label{fig:fit}
\end{figure}

We now turn to the experiments. Ref.~\onlinecite{BehBalKop07a} identified three clear peaks in the Nernst signal beyond the bulk quantum limit. In Fig.~\ref{fig:fit}B, we fit these peaks with Eq.~(\ref{eq:spec}). The index $n$ now counts the Landau levels of the surface. The quantum limit is found to be $B_1(g_{\mathrm{eff}}=0)=63$~T when the Zeeman coupling is neglected. Also shown are fits obtained for $g_{\mathrm{eff}}=8$ and $15$ with $B_1=54$~T and $43$~T, respectively. In this surface scheme, the peaks in Ref.~\onlinecite{BehBalKop07a} correspond to $n=2,3$ and $5$. The bulk quantum limit coincides with the surface $n=7$ Landau level. The $n=4$ and $6$ Landau levels correspond to fields $15.7$~T and $10.5$~T, respectively.  We reproduce in Fig.~\ref{fig:fit}A the data of Ref.~\onlinecite{BehBalKop07a} with the above features marked. Interestingly, the $n=4$ and $6$ Landau levels seem to show up in this data, in addition to the more clear peaks originally identified. The $n=6$ level is strongly shadowed by the bulk quantum limit signal, and the $n=4$ peak seems to develop for temperatures $<0.83$~K and is clearly visible, though rather broad, at $0.56$~K.

We submit that the good fit obtained here with the extended set of peaks observed in Ref.~\onlinecite{BehBalKop07a} is evidence of the surface origin of these peaks. More recently, further magnetotransport experiments by Huber~\emph{et al.}~\cite{HubNikKon08a} on bismuth nanowires in fields up to $14$~T have also clearly identified peaks caused by surface Landau level structure. The period of oscillations in nanowires is $\sim0.025$~T$^{-1}$, larger than that found here. This could be caused by the different geometry of the nanowire surfaces which can influence the size of the Fermi pockets.

We briefly discuss the effects of tilting the magnetic field on the surface states. For a field at an angle $\theta$ with the trigonal axis, the effective field entering the orbital coupling is reduced by a factor $\cos\theta$. The Zeeman coupling should also include a term $-g_\perp\mu_B\boldsymbol{\sigma}\cdot\vex B_\perp$, where $\vex B_\perp$ is the in-plane component of the field. That the orbital coupling is not affected by $\vex B_\perp$ can be seen by choosing a gauge where $\vex A=(0,Bx\cos\theta,|\vex B_\perp\times\vex r|)$ where $\vex r=(x,y)$ is the in-plane coordinate vector. 
The in-plane Zeeman coupling has the effect of shifting the momenta along $\vex B_\perp\times\hat{\vex z}$. In addition, it results in an increase in the Fermi energy and introduces an angular dependence in the spectrum thus making the Fermi surface slightly anisotropic with a remaining reflection symmetry around $\vex B_\perp$. However, unlike the $\hat{\vex z}$-axis Zeeman coupling, it does not lead to a significant change in the size of the Fermi pockets. Its strength is further diminished at small $\theta$ and therefore we do not expect a significant effect from the in-plane field on the surface states and the period of quantum oscillations, unless $g_\perp$ is anomalously large.

Shubnikov--de Haas measurements of single-crystal thin films ($\sim 10~\mu$m) of bismuth up to  a few Tesla show only the bulk carriers~\cite{YanLiuHon00a}. Why are the surface-state peaks not seen at lower fields? While we don't have a quantitative answer, we can qualitatively explain this ``low-field invisibility'' of surface states as follows. Firstly, the peaks are seen when $\omega_c\tau>1$, where $\tau$ is the relaxation time of the carriers. Since $\omega_c\propto m^{-1}$ it is smaller by roughly an order of magnitude at the surface than the bulk. The relaxation time at the surface is also expected to be shorter than the bulk. So, the surface peaks can only be seen at fields higher by more than an order of magnitude than those for the bulk, i.e. close to the bulk quantum limit. Secondly, the envelope of the oscillation peaks is expected to be different in 2d and at high fields from the usual Lifshitz-Kosevich dependence, because at high fields the chemical potential crosses only a small number of Landau levels as opposed to many in the low-field limit or in 3d~\cite{JauMarVag90a}. This results in a precipitous decrease in the amplitude as a function of $1/B$; that is,  the oscillations due to the surface diminish faster as the field is decreased relative to the bulk.

There are additional electron pockets near the $\overline{M}$ point (not shown in Fig.~\ref{fig:spec}C) which could give rise to additional oscillations. The holes are nearly compensated with the electrons. Therefore one should expect a beating with a period $\lesssim 6$ times the period of the central electron pocket. The resulting nodes in the amplitude of oscillations might offer an explanation for the suppression of the peak near $n=4$. A clear test of the surface origin of the anomalous peaks is the dependence on the thickness of the samples. One could also try contacting only the surfaces for comparison. A whole family of bismuth-based materials show similar surface states offering another venue for testing their transport signatures. This is especially illuminating in topological insulators~\cite{XiaWraQia08a} where the bulk states are gapped. It is conceivable that the anomalies are caused by a bulk reorganization of electrons~\cite{AliBal08a}. This is even more plausible if accompanied by a drop in the longitudinal resistivity. However, absent further evidence, we believe the surface theory provides a satisfactory explanation of the current data.

The authors acknowledge useful discussion with K. Behnia, B. Fauqu\'e, L. Li and P. N. Ong, and correspondence with C. Ast, H. H\"ochst, and T. Huber. This research has been supported in part by the ICMT at UIUC and the NSF Grant No. DMR0605769.
 

\end{document}